\documentclass{ifacconf}

\usepackage{graphicx}      % include this line if your document contains figures
\usepackage{natbib}        % required for bibliography
\usepackage{xcolor}
\usepackage{tikz}
\usepackage{pgfplots}
\usepackage{listings}
\usepackage{url}

\definecolor{codegreen}{rgb}{0,0.6,0}
\definecolor{codegray}{rgb}{0.5,0.5,0.5}
\definecolor{codepurple}{rgb}{0.58,0,0.82}
\definecolor{backcolour}{rgb}{0.95,0.95,0.92}
 
\lstdefinestyle{mystyle}{
    backgroundcolor=\color{backcolour},   
    commentstyle=\color{codegreen},
    keywordstyle=\color{magenta},
    numberstyle=\tiny\color{codegray},
    stringstyle=\color{codepurple},
    basicstyle=\ttfamily\footnotesize,
    breakatwhitespace=false,         
    breaklines=true,                 
    captionpos=b,                    
    keepspaces=true,                 
    numbers=left,                    
    numbersep=5pt,                  
    showspaces=false,                
    showstringspaces=false,
    showtabs=false,                  
    tabsize=2,
    escapechar=¤
}
 
\usepackage{amsmath}
\usepackage{bm}
\usepgfplotslibrary{groupplots}
\pgfplotsset{compat=newest}

\usepackage{fancyhdr}
\usepackage{nopageno}
\begin{document}
\begin{frontmatter}

\title{Developing a Hybrid Data-Driven, Mechanistic Virtual Flow Meter - a Case Study }%\thanksref{footnoteinfo}} 
%Developing a Hybrid Data-Driven, First-Principles Virtual Flow Meter - a case study
%A Hybrid Data-Driven First Principle Choke Model - a Case Study with Historical Production Data
%\thanks[footnoteinfo]{Sponsor and financial support acknowledgment
%goes here. Paper titles should be written in uppercase and lowercase
%letters, not all uppercase.}

\author[First]{M. Hotvedt} 
\author[Second]{B. Grimstad} 
\author[First]{L. Imsland}

\address[First]{Engineering Cybernetics Department, NTNU, Trondheim, Norway (e-mail: \{mathilde.hotvedt, lars.imsland\}@ntnu.no)}
\address[Second]{Solution Seeker (e-mail: bjarne.grimstad@solutionseeker.no)}

\begin{abstract}                % Abstract of not more than 250 words.
Virtual flow meters, mathematical models predicting production flow rates in petroleum assets, are useful aids in production monitoring and optimization. Mechanistic models based on first-principles are most common, however, data-driven models exploiting patterns in measurements are gaining popularity. This research investigates a hybrid modeling approach, utilizing techniques from both the aforementioned areas of expertise, to model a well production choke. The choke is represented with a simplified set of first-principle equations and a neural network to estimate the valve flow coefficient. Historical production data from the petroleum platform \textit{Edvard Grieg} is used for model validation. Additionally, a mechanistic and a data-driven model are constructed for comparison of performance. A practical framework for development of models with varying degree of hybridity and stochastic optimization of its parameters is established. Results of the hybrid model performance are promising albeit with considerable room for improvements.
\end{abstract}


\end{frontmatter}
%===============================================================================
\section{Introduction}
\thispagestyle{fancy}
\chead{\textit{© 2020 M. Hotvedt, B. Grimstad, L. Imsland. This work has been accepted to IFAC for publication under a Creative Commons Licence CC-BY-NC-ND}}
\pagenumbering{gobble}
\vspace{-0.3cm}
For a petroleum asset to succeed economically, the operators have to make crucial decisions regarding optimization of the asset. Knowledge regarding the multiphase flow rates in the asset is therefore of high importance. The flow rates may be obtained with deduction well testing, test separators and multiphase flow meters (MPFM), however, these methods are costly and MPFMs call for well intervention upon failure \citep{Marshall_well_test}. An alternative is virtual flow meters (VFM) that take advantage of measurements to describe the input-output relationship of a system with a mathematical model \citep{Toskey_VFM}. 

There are several types of VFM models. Dependent on the amount of available process data and prior knowledge of the system, the types may be placed on a scale ranging from mechanistic models (M-models) derived from first-principles, to data-driven models (DD-models), which are generic mathematical models fitted to input-output data \citep{Stosch_HM}, see Fig.~\ref{fig:scaleModels}. Often, the two extremes are called white-box and black-box models, with reference to the extent of prior knowledge about the system, for example physical interpretation of parameters and relationship between variables. The models in between are hybrid models (H-models) or gray-box models, which utilize modeling techniques from both fields and have a mixture of physical and non-physical parameters. 

In this research, an H-model of a well production choke is developed using historical production data from the petroleum platform \textit{Edvard Grieg} \citep{EG}. In addition, an M-model and a DD-model are developed for comparison of performance. A practical framework facilitating development of models with varying degree of hybridity and stochastic optimization of model parameters is constructed and conveniently enables future research into the field of hybrid modeling. Background into VFM modeling and the contributions of this research is given in Section \ref{sec:background}, the three model types of the production choke is presented in Section \ref{sec:models}, the practical framework is outlined in Section \ref{sec:framework}, the Edvard Grieg case study is presented in Section \ref{sec:caseStudy}, before simulation results and a conclusion is given in Sections \ref{sec:results} and \ref{sec:conclusion}.
\section{Background}\label{sec:background}
\subsection{Virtual flow meter modeling approaches}\label{sec:background_modeling}
\begin{figure}
\centering
\includegraphics[width=6.5cm]{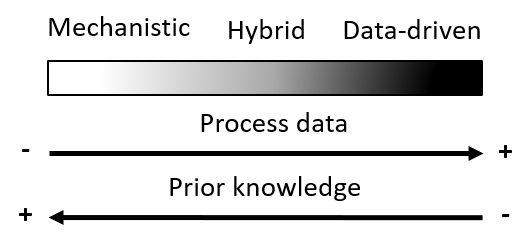}   % printed width is 8.4
\caption{Range of VFM models from mechanistic to data-driven, white-box to black-box.} 
\label{fig:scaleModels}
\end{figure}

The most common way to model VFM in today's oil and gas industry are with M-models, where some well known commercial VFM are Olga, K-Spice and FlowManager \citep{Timur_HM}. A great advantage with M-models is their way of representing prior knowledge through the use of first-principles, which leads to interpretable parameters and usually good extrapolation abilities. However, in order for M-models to be computationally feasible, model simplifications are usually a necessity and plant-model mismatch is unavoidable \citep{Solle_HM}. Additionally, in complex processes, unknown physical relations are ofttimes present and difficult to capture. VFM with DD-models have shown promising performance suitable for real-time monitoring, without the need of prior knowledge about the system \citep{ALQutami_DDM}. Further, unknown phenomena may be captured if reflected in the process measurements. However, DD-models are data hungry (Fig.~\ref{fig:scaleModels}), they struggle with extrapolation in unseen operational settings, parameters generally lack physical interpretation and incorporating process constraints may be challenging, although existent dependent on the DD-method \citep{Pitarch_HM}. Several industrial and academic M- and DD-models are reported in \citet{Mokhtari_MM, Balaji_DDM, ALQutami_DDM, Timur_HM} and references therein.

An in-between solution designed to utilize the best of both worlds are H-models. First, notice that the expression "H-models" is widely used in literature for other concepts than combinations of M- and DD-models. Further, one should differ between a hybrid model development procedure and a hybrid model in application. To clarify, most M-models use real data for parameter estimation. Thus, these models are hybrid in their development procedure, however, after development, parameters are fixed, and the model in application is an M-model. Likewise, a DD-model trained on generated data from an M-model would be hybrid in development, although not in application. Therefore, in this article, we define an H-model as follows:

\subsubsection{Definition} A hybrid model combines equations from first-principles with generic mathematical structures, both in model development and application. 

\begin{figure}
\centering
\includegraphics[width=7.0cm]{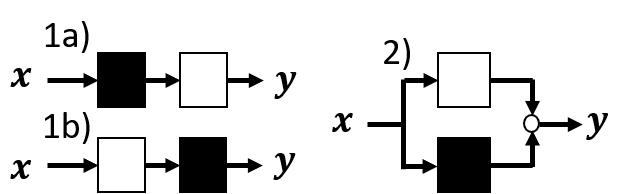}   % printed width is 8.4
\caption{Illustration of hybrid model variants. Serial H-models (type 1) and parallel H-models (type 2).} 
\label{fig:typesHM}
\end{figure}

Following the definition, an H-model is fundamentally categorized in two ways, serial or parallel, see Fig.~\ref{fig:typesHM}. Examples of serial models are online (that is, at each new state) parameter estimation with a DD-model (1a), a DD-model to capture unknown physical phenomena or modeling errors (1b) and physical equations utilized to construct specialized features as input to the DD-model, called feature engineering (1b). A parallel H-model (type 2) would be achieved if a composition of M- and DD-submodels are connected or used in an ensemble model. Naturally, combinations of the two fundamental ways will also be an H-model. 
Expectantly, compared to an M-model, the H-model should have an increased ability to capture unknown phenomena, yet have better interpretability than a DD-model through the inclusion of prior knowledge and physical parameters. Generally, the DD-part in the H-model will be smaller (in terms of number of parameters) than in a DD-model and should thus require fewer data samples to obtain a satisfactory approximation of the process \citep{Psichogios_HM}, see Fig.~\ref{fig:scaleModels}. 
\subsection{Hybrid models in literature}\label{sec:background_literature}
Some of the earliest reported H-models are within the field of chemistry \citep{Psichogios_HM, Kramer_HM}. However, H-models for VFM are rare although some examples exist in literature. For instance, \citet{Xu_HM} used feature engineering in a neural network for wet gas metering. Although feature engineering has shown to boost DD-models, choosing appropriate features is challenging \citep{RL_book}. \citet{Al-Rawahi_HM} estimated the mixture density of multiphase flow using a neural network. However, the neural network required the underlying primary measurements from a MPFM, which may not be as readily available as other measurements. Additionally, MPFM are known to require frequent calibration and may yield high measurement error in-between calibrations \citep{Falcone_MPFM}. Although not a VFM, \citet{Baraldi_HM} used an ensemble H-model to detect degradation of production choke valves. 

\subsection{Contributions}\label{sec:contributions}
The contributions of this research are two-fold:
\begin{itemize}
    \item A practical and convenient framework to facilitate development of models with varying degree of hybridity and stochastic optimization of the model parameters.
    \item A hybrid VFM model for production chokes, developed and validated utilizing real historical production data with readily available measurements such as pressures, temperatures and choke openings. 
\end{itemize}
It must be specified that the main ambition of this research has been to establish a convenient framework for development and utilization of hybrid models. In addition, this research attempts to highlight that H-models may offer advantages over M- and DD-models. Therefore, only one type H-model (type 1a) with parameter estimation using a neural network in an existing M-model has been developed. However, a notable feature with the framework is that, regardless of the hybrid model structure, the model may be trained requiring only measurements of the \textit{output} variable. 

\section{Choke models}\label{sec:models}
\begin{figure}
\centering
\includegraphics[width=5.cm]{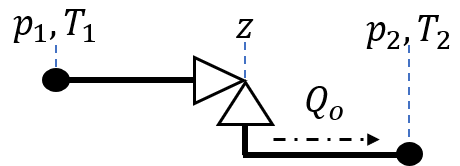}   % printed width is 8.4
\caption{Illustration of the well production choke} 
\label{fig:choke}
\end{figure}

A production choke may be illustrated as in Fig.~\ref{fig:choke}, where the volumetric oil flow rate, $Q_o$, will be estimated using nearby measurements; pressures ($p$), temperatures ($T$) and choke opening ($z$). Three model types have been developed, M-, H- and DD-model. In short notation, these are represented with $\bm{\hat{y}}_{\xi} = f_{\xi} (\bm{x}_{\xi};\bm{\theta}_{\xi})$ where $\xi \in \{m,h,dd\}$, $\bm{\hat{y}}_{\xi} = \bm{Q}_o^e$ is the estimated oil flow rate, $f_{\xi}$ are the set of model equations, $\bm{x}_{\xi}$ are the input measurements and $\bm{\theta}_{\xi}$ are learnable model parameters. In the model development, also called training procedure, an optimization algorithm finds the $\bm{\theta}_{\xi}$ that minimizes the deviation between estimated ($\bm{Q}_o^e$) and existing measurements ($\bm{Q}_o^m$) of the volumetric oil flow rate, see Section \ref{sec:framework}. The following Sections (\ref{sec:m-model}, \ref{sec:dd-model}, \ref{sec:h-model}) briefly explain the three model types and Table  \ref{tb:model_types_overview} gives an overview of inputs and parameters to the three models and highlights the difference between the M- and H-model, in this case the form of the $C_v$-curve. 

\subsection{Mechanistic model}\label{sec:m-model}
The M-model is from \citet{Kittelsen_MM}, chosen for its simplicity, and described with the equations (\ref{eq:volum_oil_flow})-(\ref{eq:xp}). More widespread choke models exist (e.g. Sachdeva, Hydro, Al-Safran, see \citet{Haug_MM} and references therein) and should be experimented with in future work. 
\begin{equation}\label{eq:volum_oil_flow}
    Q_o = \frac{w_o \dot{m}}{\rho_{o,ST}} \hspace{5mm} \left[\frac{Sm^3}{h}\right]
\end{equation}
\begin{equation}\label{eq:mass_flow}
    \dot{m} = NC_v(z)\sqrt{Y^2\rho_m(p_1 - p_2)} \hspace{5mm} \left[\frac{kg}{h}\right]
\end{equation}
\begin{equation}\label{eq:ysquared}
    Y^2 = (1-\frac{1}{3}\frac{x_{lim}}{x_{TP}})^2\frac{x_{lim}p_1}{p_1-p_2} \hspace{5mm} \left[-\right]
\end{equation}
\begin{equation}\label{eq:rho_c}
    \frac{1}{\rho_m} = \frac{w_g}{\rho_{g}} + \frac{w_o}{\rho_o} + \frac{1-w_g-w_o}{\rho_w} \hspace{5mm} \left[\frac{m^3}{kg}\right]
\end{equation}
\begin{equation}\label{eq:rho_gwh}
    \rho_{g} = \frac{M_w p_{1}}{z_{g}RT_{1}} \hspace{5mm} \left[\frac{kg}{m^3}\right]
\end{equation}
\begin{equation}
    x_{lim} = \min{(x_P, x_{TP})} \hspace{5mm} \left[-\right]
\end{equation}
\begin{equation}\label{eq:xp}
     x_P = \frac{p_1-p_2}{p_1}, x_{TP} = \frac{p_1-p_2}{p_{1}}|_{c}
\end{equation}
Two important assumptions are those of frozen flow and incompressible liquid; the mass phase fractions, $\bm{w} = [w_g, w_o, w_w]$ and liquid densities $\rho_o$ and $\rho_w$ are constant in a given operating point ($ST$ for standard conditions). For this model, the valve flow coefficient; $C_v(z)$, is determined with linear interpolation between a given set of test points, which are usually from lab-experiments with water, yet calibrated to the multiphase flow once in place. Further nomenclature may be found in \citet{Kittelsen_MM}. The learnable model parameters are chosen to be $\bm{\theta}_m = [\rho_o, \rho_w, a]$, where $a$ allows the $C_v(z)$ to be shifted; $ C_{v,new}(z)= a C_{v,old}(z)$.
\subsection{Data-driven model}\label{sec:dd-model}
The DD-model is a fully-connected, feed forward neural network (NN) with the Rectified Linear Unit (ReLU) as activation function on each layer. See e.g. \citet{Balaji_DDM} for description of neural networks. The learnable parameters are the weights and biases on each layer, $\bm{\theta}_{dd} = [\bm{W}_{dd}, \bm{b}_{dd}]$.  
\subsection{Hybrid model}\label{sec:h-model}
The H-model (type 1a, Fig.~\ref{fig:typesHM}) is represented with the same equations as for the M-model (\ref{eq:volum_oil_flow})-(\ref{eq:xp}), but with the $C_v$ obtained from a fully-connected, feed forward, NN with ReLU as activation function on each layer. The mass fractions were included as inputs to the NN in an attempt to have the $C_v$-curve reflect well-specific properties. Thus, the learnable parameters are $\bm{\theta}_h = [\rho_o, \rho_w, \bm{W}_{h}, \bm{b}_{h}]$. 
\begin{table}[h]
\caption{Overview of parameters and inputs and overview of the $C_v$-curve form}
\begin{tabular}{lllllll}
              & M-model & H-model & DD-model \\ \hline
%\# parameters & 3       &   924      &  10571        \\
 $\bm{\theta}$      & $\rho_o$, $\rho_w$, $a$     & $\rho_o$, $\rho_w$, $\bm{W}_h$, $\bm{b}_h$  & $\bm{W}_{dd}$, $\bm{b}_{dd}$      \\ 
  $\bm{x}$    &  \begin{tabular}[c]{@{}l@{}}[$p_1$, $p_2$, $T_1$, \\ $z$, $w_g$, $w_o$]\end{tabular}      & \begin{tabular}[c]{@{}l@{}}[$p_1$, $p_2$, $T_1$, \\ $z$, $w_g$, $w_o$]\end{tabular}       & \begin{tabular}[c]{@{}l@{}}[$p_1$, $p_2$,  \\ $T_1$, $T_2$, \\ $z$, $w_g$, $w_o$]\end{tabular}  \\
  $C_v(\bm{x'})$ &  \begin{tabular}[c]{@{}l@{}}Linear interpolation \\ $\bm{x'}_m = [z]$ \end{tabular} & \begin{tabular}[c]{@{}l@{}}NN \\ $\bm{x'}_h = [z, w_g, w_o]$ \end{tabular} & n.a.   \\ \hline
\end{tabular}
\label{tb:model_types_overview}
\end{table}
\section{Modeling framework}\label{sec:framework}
To easily investigate different model types, a practical framework utilizing machine learning techniques is constructed\footnote{We have utilized PyTorch, but other possibilities exist such as TensorFlow.}. The framework enables a smooth transition between training a fully M-model to a fully DD-model. It consist of several parts and will be defined in the following. 
\subsection{Defining the model}\label{sec:framework_model}
\begin{figure}
	\centering
	\includegraphics[width=6cm]{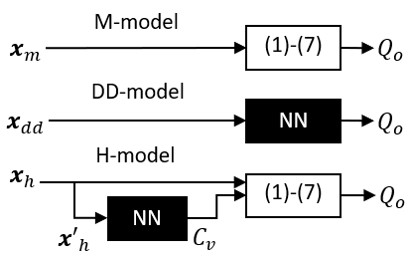}   % printed width is 8.4
	\caption{The forward pass illustrated for the three model types} 
	\label{fig:forwardPass}
\end{figure}

This part enables a convenient way to implement models with varying degree of hybridity. Firstly, the model parameters must be defined, either as single, learnable parameters, as for the physical parameters, or as NN's with weights and biases. Thus, a model will effortlessly move on the gray-scale (Fig.~\ref{fig:scaleModels}) dependent on the defined parameters. Thereafter, the forward pass, the propagation of input data through the model, will be defined as a computational graph, enabling access to the model derivatives through automatic differentiation. The forward pass for the different models is illustrated in Fig.~\ref{fig:forwardPass}. A particularly appealing property with this framework is that measurements of the $C_v$ are not required as the model is trained on the output, $Q_o$.  
\subsection{Defining the optimization problem}\label{sec:framework_opt}
Once the model is defined, a general optimization problem to find the $ \bm{\theta}_{\xi}$ that minimizes deviation between the model estimates $\bm{\hat{y}}_{\xi} =  \bm{Q}_{o,\xi}^e = f_{\xi}(\bm{x}_{\xi}; \bm{\theta}_{\xi})$ where $\xi \in \{m,h,dd\}$ and the measurements $\bm{y}=\bm{Q}_{o}^m$ may be set up as 
\begin{equation}\label{eq:optProb}
\begin{array}{llll}
    \hat{\bm{\theta}}_{\xi} &= \arg \min_{\bm{\theta}_{\xi}} J(\bm{\theta}_{\xi}, \bm{\lambda}_{\xi}) \\
    &= \arg \min_{\bm{\theta}_{\xi}}\bigg( \frac{1}{n}\sum_{i=1}^n \left(y^{(i)} - f_{\xi}(\bm{x}^{(i)}_{\xi};\bm{\theta}_{\xi} )\right)^2 \\
    &\hspace{1.6cm} + \frac{1}{n}\sum_{j=1}^p \lambda_{j,\xi} (\theta_{j,\xi} - \mu_{j,\xi})^2 \bigg)\\ 
\end{array}
\end{equation}
The first term in eq. (\ref{eq:optProb}) is the mean square error (MSE) and the second is an $\ell_2$-regularization term with regularization factors $\lambda_i$. For the physical parameters, the goal of regularization is to penalize deviation of the parameters from a prior (expected) value, $\mu_i$, and maximum a posteriori (MAP) estimation has been set up to automatically calculate the $\lambda_i$ factors, see Section \ref{sec:framework_reg}. For the NN parameters, common practice is followed and $\mu_i$ is set to zero. If $\bm{Q}_{o}^m$ are available from different measurement sources, additional MSE terms may be added and weighted according to the uncertainty in the measurement source. In this research, only measurements from a MPFM has been utilized. 

The framework solves the optimization problem in eq.~(\ref{eq:optProb}) using iterative gradient-based optimization. The update formula may be stated as follows 
\begin{equation}
    \bm{\theta}^{k+1}_{\xi} =  \bm{\theta}^{k}_{\xi} - \alpha^k \mathcal{M}(\bm{x}^{k'}_{\xi}; \bm{\theta}^{k}_{\xi},\bm{\lambda}_{\xi})
\end{equation}
where $\alpha^k$ is the learning rate (or step-size), $\bm{x}^{k'}_{\xi}$ is a subset of the data samples and $\mathcal{M}$ is the set of equations calculating the step direction. Different algorithms may be selected, such as stochastic gradient descent (SGD), Adam among others \citep{optimization_ml}. Stochastic gradient-based optimization algorithms has the advantage of being well suited for large scale models, either in terms of large data-sets or many parameters, where other optimization algorithms utilizing linesearch may be to computationally expensive \citep{Bengio_DD}. In SGD, $\mathcal{M} = \nabla_{\bm{\theta}}\tilde{J}(\bm{x}; \bm{\theta}^{k},\bm{\lambda})$, where $\nabla\tilde{J}$ may be calculated with different number of samples (batch size). Knowing which optimization algorithm yields the best result is challenging as it might be problem dependent. Therefore, the framework promote investigation of different optimization algorithms. In this research, Adam is used for all models.
\subsection{Calculation of regularization parameters}\label{sec:framework_reg}
The $\lambda_i$ regularization factors for the physical parameters may be automatically calculated through MAP estimation. If one assumes a model of the form
\begin{equation}
    y = f(x;\theta) + \epsilon \qquad \epsilon \sim \mathcal{N}(0,\sigma_{\epsilon}^2) 
\end{equation}
the MAP estimation may be set up as follows utilizing Bayes' rule, where $(X, y)$ is the collection of data points
\begin{equation}
\begin{array}{ll}
    \hat{\theta}_{MAP} &= \arg \max_{\theta} \left(\log p(y|X, \theta) + \log p(\theta)\right)
\end{array}
\end{equation}
If one additionally assumes independent Gaussian priors of the parameters $\theta_i \sim \mathcal{N}(\mu_i, \sigma_i^2)$, the MAP estimation will result in, after some rearrangements, 
\begin{equation}
\begin{array}{ll}
    \hat{\theta}_{MAP} &= \arg \min_{\theta} \bigg(\sum_{i=1}^n \left( y^{(i)} - f(x^{(i)};\theta) \right)^2 \\
    &+ \sum_{i=1}^p \frac{\sigma_{\epsilon}^2}{\sigma_i^2}(\theta_i - \mu_i)^2 \bigg)
\end{array}
\end{equation}
Dividing by $n$ and setting $\lambda_i = \sigma_{\epsilon}^2 / \sigma_i^2$, the MAP estimation will be the same as the estimate in eq. (\ref{eq:optProb}). The $\sigma_i$ may be determined based on physical bounds and if $\sigma_{\epsilon}$ is known, $\lambda_i$ is automatically calculated. In practice, $\sigma_{\epsilon}$ must be tuned, however, the number of coefficients to determine decreases.  
\section{Case study - Edvard Grieg}\label{sec:caseStudy}
Historical production data from \textit{Edvard Grieg} has been utilized in the model development procedure and to analyze performance of the models. In addition to pressures, temperatures, and choke opening (see Fig.~\ref{fig:choke}), measurements from a MPFM located upstream the choke restriction was used for training the model, keeping in mind that MPFM measurements may be faulty and require frequent calibration \citep{Falcone_MPFM}. Future work should include well-tests which in general have higher accuracy than MPFM measurements. The production data are from 10 oil wells, yielding a total of 30 models, over a period of 1248 days. Consequently, the assumption of constant physical parameters may be a rough approximation and future work should consider updating the models at certain intervals in time to account for changes in the true process.  

The data was preprocessed in two steps before performing modeling. First, the raw production data was processed by Solution Seeker's data squashing technology \citep{Grimstad2016}. The data squashing algorithm partitions the data into intervals of steady-state operation. The data in each interval is then compressed to mean values using statistics suitable for time-series data. The result is a compressed data set of steady-state operating points, suitable for steady-state modeling. In the second preprocessing step, samples considered invalid, such as samples with unrealistically large well head pressures or negative flow rates, were removed and some samples were slightly modified, for instance small negative flow rates, where measurement noise was the likely cause of error. The second step resulted in a variable number of samples per well, in the range 612-2175. Further, the mass fractions were calculated using MPFM flow rates and standard densities. In an industrial setting, the mass fractions are often calculated from sparse well-test samples, thus to mimic this setting, a mass fraction update time of 30 days was employed, using an average of the last 20 samples. 
 
The data set of each well was divided into two, training (75\%) and test (25\%), where 15\% of the latest training data was used as a validation set to decide upon the hyperparameters in the training procedure. An ambition was for the three model types to generalize well across all wells of the asset. Consequently, the same set of hyperparameters was used for a model type, instead of individual tuning of each model type for each well. However, one should expect a lower overall error by individual tuning due to dissimilar well operating conditions and variable sample numbers, and this should be considered in future work. The average root mean square error (RMSE) and average mean absolute error (MAE) of the 10 wells were monitored and the best set of hyperparameters was chosen based on the minimum obtained averages. However, if prominent overfitting occurred in a well for a set of hyperparameters, that is, if the validation error increased when the training error decreased towards the end of training, the next best set of hyperparameters was chosen. Practical recommendations from \citet{Bengio_DD} was followed in the tuning process. 

For all models, the learning rate ($\alpha$) was thoroughly experimented with as this often is the most important hyperparameter to tune \citep{Bengio_DD}. Further, for the M-model, the physical parameters had to converge within the specified bounds, thus, the number of epochs (E), that is, the number of loops through the training set, and $\sigma_{\epsilon}$ were tuned thereafter. For the H-model, both physical and NN parameters had to be found. However, E may be high and the NN architecture (width/depth) large without leading to overfitting of the NN as long as regularization of the NN parameters is applied \citep{Bengio_DD}. Hence, E was set sufficiently high and $\sigma_{\epsilon}$ adjusted for convergence of the physical parameters within bounds, the width/depth was set to 20/2 and combinations of $\alpha$ and the NN regularization factor, $\lambda_{i,nn}$, were tested. For the DD-model, the same recommendations were followed. The E was set high and combinations of $\alpha$ and $\lambda_{i,nn}$ investigated. The width/depth was set to 70/2. Lastly, the batch size (B) is often tuned independently of the other hyperparameters \citep{Bengio_DD} and was thus tuned last. Even though considerable effort was put into fair tuning of the three models, a Bayesian optimization approach will be investigated in the future to avoid (non-intentional) advantage to either model. 

An overview of the final hyperparameters are given in Table \ref{tb:hyperparameters}. Observe that the M-model required a larger $\sigma_{\epsilon}$ than the H-model for the physical parameters to converge within specified bounds, indicating that the H-model accounts for some of the measurements noise with the DD-part. Further, the best performance for the H-model was obtained with a low batch number, however, only small differences in average error lead to this choice. 
\begin{table}[h]
\caption{Overview of the final model hyperparameters}
\begin{tabular}{lllllll}
              & M-model & H-model & DD-model \\ \hline
%\# parameters & 3       &   924      &  10571        \\
 E      & 5000     & 2000      & 2000      \\
  B             & 150      & 32       & 150       \\
  $\alpha$        & 0.01      & 0.01      &0.01       \\
$\sigma_{\epsilon}$ & 25   &  10       &  -               \\
$\lambda_{i,nn}$   &   -      &   0.01      &  0.001       \\ 
width/depth  &     -       &   20/2    &   70/2    \\ \hline
\end{tabular}
\label{tb:hyperparameters}
\end{table}
\section{Simulation results}\label{sec:results}
The simulation results are shown in Fig.~\ref{fig:box_plot}, where the RMSE and MAE of the test set for the 10 wells are illustrated, and Fig.~\ref{fig:cum_plot} which is a cumulative deviation plot (CDP) \citep{cum_dev_plot} indicating the accuracy of the developed VFM models, that is, how many of the test points fall within a certain deviation from the measurement. 
\begin{figure}[ht]
\centering
\resizebox{0.8\columnwidth}{!}{%
% This file was created by matplotlib2tikz v0.7.4.
\begin{tikzpicture}

\begin{axis}[
legend cell align={left},
legend style={at={(0.03,0.97)}, anchor=north west, draw=white!80.0!black, font=\large},
tick align=outside,
tick pos=left,
x grid style={white!69.01960784313725!black},
xmajorgrids,
xmin=-0.425, xmax=8.5,
xtick style={color=black},
xtick={1.5,4.5,7.5},
xticklabels={M-model,H-model,DD-model},
ticklabel style = {font=\normalsize},
y grid style={white!69.01960784313725!black},
ymajorgrids,
ymin=-1.13184008758051, ymax=45.7686418391908,
ytick style={color=black}
]
\addplot [very thick, blue, forget plot]
table {%
0.7 7.2632708639451
1.3 7.2632708639451
1.3 16.2765397755388
0.7 16.2765397755388
0.7 7.2632708639451
};
\addplot [very thick, blue, forget plot]
table {%
1 7.2632708639451
1 4.1719702667517
};
\addplot [very thick, blue, forget plot]
table {%
1 16.2765397755388
1 23.2174385435773
};
\addplot [very thick, blue, forget plot]
table {%
0.85 4.1719702667517
1.15 4.1719702667517
};
\addplot [very thick, blue, forget plot]
table {%
0.85 23.2174385435773
1.15 23.2174385435773
};
\addplot [very thick, red, forget plot]
table {%
1.7 6.38408578835196
2.3 6.38408578835196
2.3 12.0662465686884
1.7 12.0662465686884
1.7 6.38408578835196
};
\addplot [very thick, red, forget plot]
table {%
2 6.38408578835196
2 3.3056620687072
};
\addplot [very thick, red, forget plot]
table {%
2 12.0662465686884
2 15.2524607119353
};
\addplot [very thick, red, forget plot]
table {%
1.85 3.3056620687072
2.15 3.3056620687072
};
\addplot [very thick, red, forget plot]
table {%
1.85 15.2524607119353
2.15 15.2524607119353
};
\addplot [very thick, blue, forget plot]
table {%
3.7 6.820592502025
4.3 6.820592502025
4.3 15.935392403221
3.7 15.935392403221
3.7 6.820592502025
};
\addplot [very thick, blue, forget plot]
table {%
4 6.820592502025
4 1.70285833986579
};
\addplot [very thick, blue, forget plot]
table {%
4 15.935392403221
4 25.4923785707117
};
\addplot [very thick, blue, forget plot]
table {%
3.85 1.70285833986579
4.15 1.70285833986579
};
\addplot [very thick, blue, forget plot]
table {%
3.85 25.4923785707117
4.15 25.4923785707117
};
\addplot [very thick, red, forget plot]
table {%
4.7 5.75009085281551
5.3 5.75009085281551
5.3 13.2566900640804
4.7 13.2566900640804
4.7 5.75009085281551
};
\addplot [very thick, red, forget plot]
table {%
5 5.75009085281551
5 1.37426997262078
};
\addplot [very thick, red, forget plot]
table {%
5 13.2566900640804
5 20.0172127258531
};
\addplot [very thick, red, forget plot]
table {%
4.85 1.37426997262078
5.15 1.37426997262078
};
\addplot [very thick, red, forget plot]
table {%
4.85 20.0172127258531
5.15 20.0172127258531
};
\addplot [very thick, blue, forget plot]
table {%
6.7 7.27068847426103
7.3 7.27068847426103
7.3 11.5197326329913
6.7 11.5197326329913
6.7 7.27068847426103
};
\addplot [very thick, blue, forget plot]
table {%
7 7.27068847426103
7 3.58741473672424
};
\addplot [very thick, blue, forget plot]
table {%
7 11.5197326329913
7 18.2606039866011
};
\addplot [very thick, blue, forget plot]
table {%
6.85 3.58741473672424
7.15 3.58741473672424
};
\addplot [very thick, blue, forget plot]
table {%
6.85 18.2606039866011
7.15 18.2606039866011
};
\addplot [very thick, blue, mark=*, mark size=3, mark options={solid,fill opacity=0,draw=black}, only marks, forget plot]
table {%
7 43.6368017516103
};
\addplot [very thick, red, forget plot]
table {%
7.7 5.53363731581087
8.3 5.53363731581087
8.3 9.31594237968742
7.7 9.31594237968742
7.7 5.53363731581087
};
\addplot [very thick, red, forget plot]
table {%
8 5.53363731581087
8 3.17379639781038
};
\addplot [very thick, red, forget plot]
table {%
8 9.31594237968742
8 12.5601855595907
};
\addplot [very thick, red, forget plot]
table {%
7.85 3.17379639781038
8.15 3.17379639781038
};
\addplot [very thick, red, forget plot]
table {%
7.85 12.5601855595907
8.15 12.5601855595907
};
\addplot [very thick, red, mark=*, mark size=3, mark options={solid,fill opacity=0,draw=black}, only marks, forget plot]
table {%
8 38.3985970419903
};
\addplot [very thick, blue]
table {%
%0 1
1 1
};
\addlegendentry{RMSE}
\addplot [very thick, red]
table {%
%0 1
1 1
};
\addlegendentry{MAE}
\addplot [very thick, black, forget plot]
table {%
0.7 11.1331106343197
1.3 11.1331106343197
};
\addplot [very thick, black, forget plot]
table {%
1.7 9.28101996994242
2.3 9.28101996994242
};
\addplot [very thick, black, forget plot]
table {%
3.7 13.0620943548905
4.3 13.0620943548905
};
\addplot [very thick, black, forget plot]
table {%
4.7 8.86555561407825
5.3 8.86555561407825
};
\addplot [very thick, black, forget plot]
table {%
6.7 8.10513702183687
7.3 8.10513702183687
};
\addplot [very thick, black, forget plot]
table {%
7.7 7.21786034085729
8.3 7.21786034085729
};
\end{axis}

\end{tikzpicture}%
}
\caption{Boxplot overview with median for the three model types across the 10 wells.}
\label{fig:box_plot}
\end{figure}
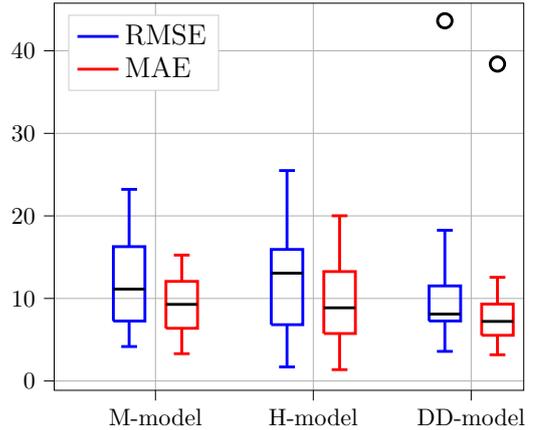
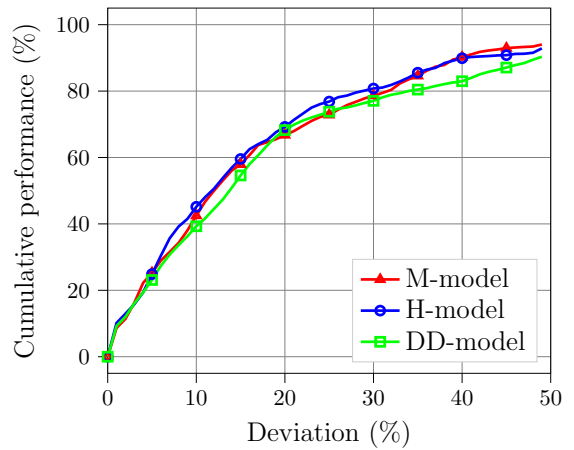
\begin{figure}[ht]
\centering
\resizebox{0.85\columnwidth}{!}{%
% This file was created by matplotlib2tikz v0.7.4.
\begin{tikzpicture}

\definecolor{color0}{rgb}{0.12156862745098,0.466666666666667,0.705882352941177}
\definecolor{color1}{rgb}{1,0.498039215686275,0.0549019607843137}
\definecolor{color2}{rgb}{0.172549019607843,0.627450980392157,0.172549019607843}

\begin{axis}[
legend cell align={left},
legend style={at={(0.97,0.03)}, anchor=south east, draw=white!80.0!black, font=\large},
tick align=outside,
tick pos=left,
x grid style={white!50.19607843137255!black},
xlabel={Deviation (\%)},
xlabel style = {font=\large},
xmajorgrids,
xmin=0, xmax=50,
xtick style={color=black},
y grid style={white!50.19607843137255!black},
ylabel={Cumulative performance (\%)},
ylabel style = {font=\large},
ymajorgrids,
ymin=-5, ymax=105,
ytick style={color=black}
]
\addplot [very thick, red, mark=triangle,  mark repeat=5]
table {%
0 0
1 8.6737400530504
2 11.3793103448276
3 16.2068965517241
4 22.2281167108753
5 25.2785145888594
6 28.9655172413793
7 31.5649867374005
8 34.2705570291777
9 38.0371352785146
10 42.4933687002653
11 46.710875331565
12 50
13 53.1034482758621
14 56.0477453580902
15 57.9310344827586
16 60.7161803713528
17 63.7665782493369
18 64.6153846153846
19 65.5172413793103
20 66.763925729443
21 67.9310344827586
22 69.4960212201592
23 71.0079575596817
24 72.2281167108753
25 72.8381962864721
26 74.3766578249337
27 75.7559681697613
28 76.763925729443
29 77.7984084880637
30 78.6472148541114
31 79.4429708222812
32 80.3713527851459
33 82.3342175066313
34 83.5543766578249
35 84.5888594164456
36 86.1273209549072
37 87.3474801061008
38 87.8779840848806
39 89.5225464190981
40 90.2122015915119
41 91.0344827586207
42 91.8037135278515
43 92.2281167108753
44 92.5464190981432
45 92.9442970822281
46 93.1299734748011
47 93.289124668435
48 93.42175066313
49 94.0318302387268
};
\addlegendentry{M-model}
\addplot [very thick, blue, mark=o, mark repeat=5]
table {%
0 0
1 10.159151193634
2 12.9442970822281
3 15.7824933687003
4 19.3633952254642
5 24.8806366047745
6 30.4244031830239
7 35.6763925729443
8 39.2042440318302
9 41.5119363395225
10 45.1458885941645
11 48.0636604774536
12 50.6100795755968
13 53.8461538461538
14 56.9230769230769
15 59.5490716180371
16 62.4933687002653
17 64.0318302387268
18 65.3050397877984
19 67.6392572944297
20 69.2042440318302
21 71.2997347480106
22 73.1564986737401
23 75.0132625994695
24 76.1007957559682
25 76.8435013262599
26 78.1432360742706
27 78.6472148541114
28 79.5490716180371
29 80.1591511936339
30 80.7692307692308
31 81.0610079575597
32 81.8567639257295
33 82.9177718832891
34 84.1379310344828
35 85.5437665782493
36 86.4456233421751
37 87.0026525198939
38 88.4350132625995
39 89.1777188328912
40 89.893899204244
41 90.2652519893899
42 90.4244031830239
43 90.5570291777188
44 90.7427055702918
45 90.8222811671088
46 91.1936339522546
47 91.2466843501326
48 91.5384615384615
49 92.8647214854111
};
\addlegendentry{H-model}
\addplot [very thick, green, mark=square, mark repeat=5]
table {%
0 0
1 9.25729442970822
2 12.0159151193634
3 16.0212201591512
4 19.4164456233422
5 23.1034482758621
6 27.3740053050398
7 30.7692307692308
8 33.6074270557029
9 36.2864721485411
10 39.2838196286472
11 41.6710875331565
12 44.5358090185676
13 47.3740053050398
14 50.8753315649867
15 54.5888594164456
16 57.9840848806366
17 60.6896551724138
18 63.6339522546419
19 66.2599469496021
20 68.2493368700265
21 69.867374005305
22 71.0875331564987
23 72.0424403183024
24 72.8912466843501
25 73.7931034482759
26 74.6684350132626
27 75.0663129973475
28 75.7029177718833
29 76.4456233421751
30 77.0557029177719
31 78.0901856763926
32 78.8859416445623
33 79.3633952254642
34 80.053050397878
35 80.4509283819629
36 80.8753315649867
37 81.5384615384615
38 82.0954907161804
39 82.5994694960212
40 82.9708222811671
41 83.9522546419098
42 85.0397877984085
43 85.8090185676393
44 86.3925729442971
45 87.0822281167109
46 87.8249336870027
47 88.4350132625995
48 89.4429708222812
49 90.3448275862069
};
\addlegendentry{DD-model}
\end{axis}

\end{tikzpicture}%
}
\caption{Cumulative deviation plot of all test samples across the 10 wells.}
\label{fig:cum_plot}
\end{figure}
There are several interesting observations to be made from the results. Firstly, notice the extreme outlier that is present in the DD-model performance in Fig.~\ref{fig:box_plot}. The outlier is caused by one of the wells which had an operational setting very different from the setting in the training set. As mentioned in Section \ref{sec:background}, DD-models may struggle with extrapolation in unseen operational settings which may explain the outlier. In that case, the results indicate that the H-model has preserved some of the extrapolation power of the M-model which do not have the extreme outlier. However, despite the H-model obtaining the lowest errors for some of the wells, the M-model performs better than the H-model due to less spread between the quartiles. Now, the only difference between the two models is the form of the $C_v$-curve. This indicates that our prior belief of the $C_v$-curve in the M-model was good and that no flexibility was added to the H-model by having the $C_v$ as an NN. Naturally, these results are preliminary and further investigations are necessary. In particular, different H-model variants may better leverage the advantages of both M- and DD-models.  

Generally, the results show a higher error than expected. Other studies have reported almost 90\% performance for 20\% deviation in the CDP (e.g. \citet{ALQutami_DDM}), whereas in this paper about 70\% performance for 20\% deviation is achieved. There may be several causes for the large error. Firstly, preprocessing of the data could be improved by for instance further outlier removals. In addition, MFPM measurements was used for mass fraction calculation and in training despite a possibility of being faulty in between calibrations. Further, the mass fractions was updated every 30 days to mimic an industrial setting, however, in training, continuous mass fraction updates could be utilized. Hence, future work should include measurement sources with higher accuracy, such as well-tests, and analyze performance with continuous mass fraction updates. Secondly, the model types were generalized across all wells and improved results would most likely be achieved with individual tuning. Further, utilization of a more accurate mechanistic choke model or optimization of additional physical parameters may decrease the error in the M- and H-model. Thirdly, the number of days for which the models are used in prediction should be taken into account. For some of the wells, the test set covered more than 200 days, whereupon the true process could have changed significantly and the models lack validity. Future work should consider online training of the models at regular intervals in time. Nonetheless, the main goal of this research was not to find exceptional models, but to illustrate that an H-model may offer advantages over M- and DD-models and to establish a framework for convenient future research. 
\section{Conclusion}\label{sec:conclusion}
Results in Section \ref{sec:results} indicate that hybrid modeling is promising and may offer advantages over both mechanistic and data-driven modeling. However, results are preliminary and there is considerable room for improvements. Future work should put more effort into preprocessing of the data set, analysis of the mass fraction calculation influence on performance and inclusion of well tests in training and validation. Further, individual tuning of each well should be investigated and the models should be tested with different data sets, for instance from other petroleum assets. Last but not least, future work should explore different hybrid model variants, for which the presented framework is convenient and highly suitable. 
\begin{ack}
This research is a part of BRU21 - NTNU Research and Innovation Program on Digital and Automation Solutions for the Oil and Gas Industry (\url{www.ntnu.edu/bru21}) and supported by Lundin Norway AS.    
\end{ack}

\bibliography{ms}             

%\appendix
%#\section{A summary of Latin grammar}    % Each appendix must have a short title.
%\section{Some Latin vocabulary}              % Sections and subsections are supported  
                                                                         % in the appendices.
\end{document}